# Gauge fields in dissipation processes

A.N.Yakunin[*]

*Karpov Institute of Physical Chemistry, 103064 Moscow, Russia*

**Abstract:** An technique is extended to estimate some critical exponents without using the expansion over the coupling constant. The data obtained is in a agreement with those found by help of the 2D Onsager method or with recent 3D results. In the second case the exponent of the correlation radius can be expressed in terms of the fine structure constant. The polymer plastic deformation process is regarded as the energy dissipation one due to the spontaneous symmetry breaking controlled by gauge fields. A comparison of the theory with other ones and different experimental data is carried out.

---

[*] E-mail: yakunin@cc.nifhi.ac.ru

# 1 Introduction

The critical phenomena and the second order phase transitions are referred to fundamental problems of modern physics. Also, they play an important role for applied science, especially upon studying (or even upon obtaining) new functional materials. However, up to now there exist some problems preventing a more wide application of received results [1-10]. Firstly, all mathematical approaches estimating, for example, the critical exponents are complex enough [1-7]. In this connection several attempts of different authors [8-10] to use more simple techniques and explanations are well-known. Their books are scientific bestsellers and almost always are often republished. But as before these approaches remain complicated for understanding. As a result it takes additional efforts from scientists and permanent investigations of specific questions [1-10]. The aim of the paper is to extend a technique suggested recently by the author [11]. The reader who knows mathematics and physics syllabi in frameworks of the first three years of an university is capable of reproducing all results. Secondly, apart education goals there are important scientific ones. Since the Wilson method [2] bases on the expansion over the small parameter, the coupling constant, it would be useful to estimate critical exponents using no additional small parameters besides "natural" - $\mu \sim 1/N$. Here, in the so-called excluded volume problem of polymer statistics, $\mu$ is the ratio of the chemical potential of a chain monomer to temperature, $T$, which is expressed in energy units, or, in other problems, the reduced deviation from the critical point [8]; $N$ is the polymerization degree or the length of a self-avoiding walk expressed in units of the lattice constant. An other aspect of the problem contains in the finite length of real polymer chains [11]. Here, it should be recalled that P.-G. de Gennes considered a polymer chain with volume interactions in the absence of an external field [3]. It means that the volume fraction of the chain ends, $N^{-\nu D}$, tended to zero as $N \to \infty$, $\nu$ was the critical exponent of the correlation radius, $D$ was the space dimension [8]. Exactly, the chain ends define the external field in this theory [4]. At last, it is possible the reader will be able to see nature of the critical phenomena otherwise. Then it may be a good stimulus for his or her in order to fulfill new original investigations and to obtain important results.

The structure of the paper is following. To confirm the validity of the technique [11] the Fisher method [12] will be modified for the two-dimensional (2D) case. The solution obtained agrees with well-known Onsager's results [9, 13]. Then the 3D case of a multicomponent ordering field will be considered. The found estimations of the critical exponents are in a good accordance with recent data [7, 14, 15] (see also refs in them). The universality of the critical exponents will be discussed in connection with a possibility of the expression of the correlation radius exponent in terms of the fine structure constant. The subsequent study is dealing with the relation of the 2D ordering parameter to the 3D one at critical point. Far from the point the 3D ordering parameter proves to be a bridge between micro and macroscopic plastic deformations of semicrystalline polymers. The latter can be described in terms of thermodynamics of irreversible processes [16]. Finally, some new solutions of old problems such as tasks of the percolation theory in gels [8, 17] will be suggested to illustrate possibilities of the technique.

## 2 RESULTS AND DISCUSSION

Following Fisher [12] let $\Pi_N$ be the total number of self-avoiding walks of $N$ steps over a simple square lattice and let $\Pi_N p_N(r)$ be the number terminating at $r$. Then

$$\sum_r p_N(r) = 1.$$

Defining the generating function by the series:

$$P(\mu, r) = \sum_{N=1}^{\infty} p_N(r) \exp(-\mu N)$$

one can obtain assuming that $\mu > 0$

$$\sum_r P(\mu, r) = 1/(exp(\mu)-1). \tag{1}$$

Thus, the sum diverges as $\mu^{-1}$ when $\mu \to 0$. One may usually observe the analogous behavior for the order parameter about the point of the second order phase transition or about critical points. Let us define the critical point as $\mu_{cr}=0$.

The main aim is to show that in the 2D case as well as in the 3D one [11] the Fishers distribution is only one factor in the expression of a full probability. The second factor can be accounted for the way we will demonstrate below.

2 1 The computation of $p_N(r)$.

To obtain $p_N(r)$, the spherical average of $p_N(\mathbf{r})$, we may invert the generation function with the aid of Cauchy's theorem [12]. Thus, with $z = e^{-\mu}$

$$p_N(r) = 1/(2\pi i) \oint dz\, (z^{-N}/z)\, P(\mu, r) = 1/(-2\pi i) \int_{c-i\pi}^{c+i\pi} d\mu\, e^{\mu N} P(\mu, r)$$

where $c$ is chosen larger than the real part of any singularity of $P(\mu, r)$. Defining $P(\mu, r)$ as

$$P(\mu, r) = P_{cr}(r)\, e^{-ikr}\, P`(\mu, ikr) \tag{2}$$

where $P`(0,0) = 1$ and

$$P_{cr}(r) \approx S_D^{-1} k_0^{1/\nu}/r^{D-(1/\nu)}, \tag{3}$$

one can see that this is in accordance with (1), since [12]

$$S_D \int_0^\infty P_{cr}(r)\, e^{-kr}\, r^{D-1}\, dr = \int_0^\infty x^{1/\nu - 1}\, e^{-x}\, dx = \Gamma(1/\nu)/\mu$$

if $x = kr$ and

$$k \approx k_0\, \mu^\nu. \tag{4}$$

$S_D = 2\pi^{D/2}/\Gamma(D/2)$ is the surface area of the D-dimensional sphere of unit radius. Substituting with (2) to (4) yields ($r, N \to \infty$)

$$p_N(r) \cong A r^{-D} N^{-1} (r/R)^{1/\nu} \int_{c'-i\infty}^{c'+i\infty} \exp(E - (r/R) E^{\nu}) P`(\mu, ikr) \, dE \quad (5)$$

where $A^{-1} = -2\pi i \, S_D$ is a constant, the variable $E = \mu N$, and the scaling factor is ($N \to \infty$)

$$R \sim k_0^{-1} N^{\nu}.$$

To integrate (5) by the method of steepest descents and to obtain the real probability density we write the "phase" factor $P`(\mu, ikr)$ (if the exponent of the correlation radius $\nu = 1$) as follows [18]

$$P`(\mu, ikr) = \exp(- (2/\pi) i E \log |E|).$$

If $\nu \neq 1$ then one can use the Fisher method [12]. Let us put $-i\varepsilon$ for $E$ and find ($\nu = 1$)

$$p_N(r) \cong (-i) A r^{-D} N^{-1} (r/R)^{1/\nu} \int_{-\infty}^{+\infty} \exp(-i\varepsilon + (r/R)\varepsilon + (2/\pi)\varepsilon \log|\varepsilon|) \, d\varepsilon. \quad (6)$$

The equation for the saddle point $\varepsilon = \varepsilon_0$ of the integrand in (6) is then

$$-i + x + (2/\pi) \log (\varepsilon_0 e) = 0$$

which yields

$$\varepsilon_0 = i \exp(- (\pi/2) (r/R) - 1).$$

Expanding the exponent about the saddle point for $r \to \infty$ leads to

$$p_N(r) \cong B/R^D (r/R)^{\varphi} \exp(-(r/R)) \quad (7)$$

where

$$B^{-1} = (\pi/2) e^{1/2} S_D, \; R = (4/\pi) k_0^{-1} N, \; \varphi = 1/\nu - D. \quad (8)$$

Thus, we obtain famous Onsager's result [13] by the Fisher method of polymer statistics [12]. To see this more clear let us go to the next point.

2 2 Calculation of the critical exponent of the "heat capacity"; an thermodynamic consideration.

The logarithm of the function (7) is proportional to $\mu N$ in accordance with the way of the definition of the function. It should be underlined that in practice we study the system with $N = const$, but not with $\mu = const$. This results in the "renormalization" of the chemical potential (see [8] as an example)

$$\mu \sim N^{\alpha-1} \qquad (9)$$

where the critical exponent, $\alpha$, may be expressed as [2, 8-10]

$$\alpha = 2 - \nu D. \qquad (10)$$

Comparing (7) with $\mu N \sim N^{\alpha}$ (9) we find for $r = C R$ ($C = const >> 1$) that

$$\alpha = 0$$

in agreement with (10).

This Onsager's result may be received by other methods [5, 9].

2 3 Definition of Green function.

Let us write the equation for the Green function of a polymer chain to which an external uniform mechanical field is applied [1, 8, 19-21]:

$$((a^2/2D)\nabla^2 + fx/T)\psi = \partial\psi/\partial N. \qquad (11)$$

Such form of the equation (11) is due to an extension of the polymer chain, since if $\nu = 1$ the chain length $\sim aN$ is equal to the end-to-end distance, $a$ is the monomer diameter (the lattice constant). Consequently, some applied force $f$ is necessary to keep the given conformation. Substituting

$$X = (fx/T + \mu)(2D)^{1/3} / (fa/T)^{2/3} \qquad (12)$$

and $g(X) = exp(-\mu N)\psi(x)$ one can obtain

$$\nabla^2 g + g X = 0.$$

The solution of this equation at $X \rightarrow -\infty$ is [19]:

$$g \sim |X|^{-1/4} \exp(-2|X|^{3/2}/3). \tag{13}$$

At the critical point ($\mu_{cr}=0$) we find from (12) that $X_{cr} \sim (fa/T)^{1/3}x/a$. Therefore, we define the correlation radius as $R = a\,(fa/T)^{-1/3}$. Since the scale parameter is single for this problem then, using (8), we may receive $fa/T \sim N^{-3\nu}$ and find $X_{cr} \sim N^{-\nu}x/a$.

The function $g$ at $x < 0$ can be regarded as a probability no penetration for the chain over the potential barrier, $-fx/T$. At the critical point ($\mu_{cr}=0$) the factors such as $|X|^{-1/4}$ in (13) and $(r/R)^\varphi$ in (7) will be play a crucial role.

2 4 Estimation of the critical exponent evaluating the total number of self-avoiding walks.

If we define the square of the ordering parameter about the critical point as ($\mu_{cr}=0$, where one may take $R \sim \xi \to \infty$)

$$W^2 = p_N(r)\,R^D\,N^{-1}/g\,(X_{cr}) \sim N^{-1-\varphi\nu-\nu/4} \sim N^{-2\beta} \tag{14}$$

then we can obtain the critical exponent, $\beta$:

$$2\beta = \nu/4 \tag{15}$$

Comparing the definition with the conventional expression [2, 8-10]

$$2\beta = \nu\,(D - 2 + \eta) \tag{16}$$

one may see ($D = 2$) that

$$\eta = 1/4. \tag{17}$$

Finally, we can find the exponent, $\gamma$, since [2, 8-10]

$$\gamma = \nu\,(2 - \eta). \tag{18}$$

Substituting (17) for $\eta$ in (18) we obtain ($\nu = 1$)

$$\gamma = 7/4$$

that agrees with the same value received by help of Onsager's theory [9, 22].

It should be noted that we define the ordering parameter (14) at the critical point ($\mu_{cr}=0$) as the ratio of the two probabilities, $p_N(r)R^D/N$ and $g(X_{cr})$, that is, $W^2$ is an condition probability as well as in the 3D case [11].

We have just modified the Fisher method. We ought only to underline that $g(X)$, by its definition [1, 8, 19-21], satisfies the condition to locate the different monomers in different sites of the lattice. Let us write the values of $\beta = 0.125$ using (15).

2 5 Multicomponent ordering fields in 3D space.

Following Edwards [1] the Green function of a polymer chain with the excluded volume can be found as a solution of the diffusion equation (compare with eq. (11))

$$\partial\psi/\partial N = ((a^2/2D)\nabla^2 - \varphi(r)/T)\psi = \hat{H}\psi \qquad (19)$$

with the boundary condition $\psi(r, r`) = \delta(r - r`) a^3$ at $N = 0$; $\varphi(r)$ is the potential describing the excluded volume parameter, $V$,

$$V = \int (1 - exp(-\varphi(r)/T)) d^3 r.$$

Let $\psi(r)$ be an eigen function of the operator $\hat{H}(r)$. Let us assume also that there exists an uniform field $H$ with a vector potential $A(r)$. While the symmetry of the operator $\hat{H}(r)$ can be broken due to the vector potential is not the periodic function of coordinates, a physical translation symmetry of the polymer system does not change, as the lattice over that the macromolecule makes random walks is periodic. Further we will use the well-known solution about the symmetry of electron states in the lattice with a magnetic field [23].

Let us choose the gauge for the vector potential of the uniform field as follows

$$A = [H\ r]/2.$$

Upon a translation $r \rightarrow r + L$ (where $L$ is one of the lattice periods) $\psi(r) \rightarrow \psi(r + L)$. The latter is the eigen function of the $\hat{H}(r + L)$ which does not coincide with $\hat{H}(r)$ due to substituting

$$A(r + L) = A(r) + [H\,L]/2$$

for $A(r)$. To return to the initial $\hat{H}(r)$ the following gauge transformations are necessary

$$A(r + L) \rightarrow A(r + L) + \nabla f, f = -[H\,L]\,r/2.$$

The wave function $\psi(r) \sim e^{ikr}$ is the eigen function of the operator $\hat{H}(r)$ (if $\varphi(r) = 0$ in (19)) as it may be seen from the solution the equation

$$-i\nabla \psi(r) = k\,\psi(r).$$

Under the gauge transformations the wave function must be multiplied by $e^{icf}$ where $c = const$:

$$(-i\nabla - cA(r))\psi \rightarrow$$

$$(-i\nabla - cA(r) - c\nabla f)e^{icf}\psi = -i\nabla(e^{icf}\psi) - e^{icf}\psi(cA(r) + c\nabla f) = e^{icf}(-i\nabla - cA(r))\psi.$$

Let the translation operator $\hat{S}_L$ be the result of all mentioned above actions:

$$\hat{S}_L\,\psi(r) = \psi(r + L)\,e^{i[h\,L]\,r/2}$$

where $h = cH$. One can obtain

$$\hat{S}_L\,\hat{S}_{L`} = \hat{S}_{L+L`}\,g(L, L`),$$

where

$$g(L, L`) = e^{-ih[L\,L`]/2} \qquad (20)$$

and

$$\hat{S}_L\,\hat{S}_{L`} = \hat{S}_{L`}\,\hat{S}_L\,g^2(L, L`).$$

If the field $h$ will be

$$h = (4\pi/p)\,v^{-1}\,a_3 \qquad (21)$$

where p is a simple number, $a_3$ is one of the lattice periods, and $v$ is the volume of the elementary unit ($v = a^3$ for the primitive cubic lattice which we will use for the sake of simplicity), then for the translations

$$\mathbf{L} = a\,(m\,\mathbf{i} + p\,n\,\mathbf{j} + l\,\mathbf{k})$$

where $\mathbf{i}, \mathbf{j}, \mathbf{k}$ are the unit vectors of the orthogonal basis, $m$, $n$, and $l$ are the whole numbers, $g(\mathbf{L}, \mathbf{L'}) \equiv 1$ in (20) and $\hat{S}_L\,\hat{S}_{L'} = \hat{S}_{L'}\cdot\hat{S}_L$.

Returning to estimations of the critical exponents we have found [11] that the wave vector $q$

$$qa \sim N^{0.5 - v} \tag{22}$$

gives the main contribution to the integral for receiving the pair correlation function. The wave vector $q$ has a very important meaning for polymer statistics, since it is connected with the fluctuation interactions of the chain ends, and, consequently, with the field $\mathbf{h}$ corresponding to the chain ends [3, 4, 8].

Let $h$ designate $\mathbf{hk}$ in (21) and let us choose $N = \exp(1/(ha^2))$. Then $N < N_{cr}$ at $h > h_{cr}$ where $N_{cr}$ is a value of the polymerisation degree at which the transition from a swelling coil to ideal ($h \to h_{cr} > 0$) can be revealed [11]. One may find

$$\log(N/N_{cr}) \approx -\Delta h/(h_{cr}a)^2 \tag{23}$$

where $\Delta h = h - h_{cr}$.

Let us assume

$$q/q_{cr} - 1 = h/h_{cr} - 1. \tag{24}$$

Using (23) and expanding the exponential form of (22) yields

$$q/q_{cr} - 1 = \exp((0.5 - v)\log(N/N_{cr})) - 1 \approx (v - 0.5)\,\Delta h/(h_{cr}a)^2. \tag{25}$$

From (24), (25) and (21) we write

$$v = 0.5 + 4\pi/p = 0.5 + 2\pi\,(n + 2)/p \tag{26}$$

where $n$ is the number of components of an ordering field, $p =137$ to obtain the best agreement with the estimation $v \approx 1 – 6^{-1/2}$ at $n = 0$ [11]. We have used the expression $v - 0.5 \sim n + 2$ by the Wilson ε-expansion [2, 10].

The formula (26) can be rewritten in other way supposing $2\pi (n + 2)/p$ by a small parameter

$$v = 1 - 0.5 (1 + 2\pi (n + 2)/p)^{-1} + \pi (n + 2)/p. \quad (27)$$

The expression (27) seems to agree with the recent results [7, 14, 15] (see refs in them).

Let us estimate the exponent $\eta$ characterizing the field anomalous dimension by the Wilson method [2, 10] together with the other technique [11]. The main assumption is to regard this exponent as a constant. The condition approximately holds in the every order of the ε-expansion [2] if $D=3$. Thus, one can assume

$$\eta \sim x\, \varepsilon^3 \quad (28)$$

where $\varepsilon = 4 – D$, and $\eta = x = 1/32 = 0.03125$ ($D=3$), since $\eta = 0.25$ if $D=2$ (17). An other estimation can be made using the expression

$$\eta \sim y\, \varepsilon^2 + x\, \varepsilon^3 \quad (29)$$

and the value $\eta \sim y \approx 0.0132$ for $D=3$ [11]. Again if $D=2$ from (29) $x = (0.25 – 0.0132 * 4)/8 = 0.02465$, and for $D=3$ $\eta \approx x + y = 0.03785$. From the obtained estimations (28) and (29) we find

$$\eta = 0.035 \pm 0.004 \quad (30)$$

and the value with its error range overlaps practically all recent data [7, 14, 15]. A small discrepancy exists only for $n = 0$ [7, 14], the precise experimental results seem to be capable of solving the future problem.

2 6 The universality of scaling laws.

If the fine structure constant

$$\alpha_{fs} = e^2/\hbar c \qquad (31)$$

(where $e$ is the electron charge, $h = \hbar/2\pi$ is the Planck constant, $c$ is the light velocity) will be approximately correspond to $p^{-1} = 1/137$ then the universality of the critical laws should not be explained – everything is clear.

We think that $\eta$ in (30) may be connected with the gauge constant $g = \eta^{-1/2}$ of a field, for example, $sin\theta_W = e/g \approx 0.457$ if $\theta_W$ is the Weinberg angle in the theory of electroweak interactions. But those will be other researches.

The rest exponents are expressed in terms of the similarity laws (10), (16), (18) [2, 7-10].

2 7 Energy dissipation during plastic deformation of semicrystalline polymers: microscopic and macroscopic considerations.

The cross-section of the melt-crystallized polymer sample will decrease dramatic, that is, the neck will form, as the force applied to the sample in the drawing direction exceeds some define limit (yielding). The phenomenon of the transition from the initial isotropic state to the final oriented one can be called for stationary cases by the neck formation. The part of literature data dealing with the stationary necking process has critically discussed by the author [21]. Non stationary behaviour presents a separate interest [24-27] and is beyond scope of the paper.

The subsequent drawing occurs through the neck propagation along the sample length in the elongation direction. It should be noted that the neck draw ratio equal to the ratio of the final sample length to its initial, $\lambda_n$, falls with the molecular weight increase following the power law with the exponent equal to $-0.3$ [21]. It means that at some critical molecular weight of polymer (or at the polymerization degree $N_{cr}$) $\lambda_n = 1$, no plastic deformation can be observed, and the sample must remain isotropic. A similar behaviour one may receive, for example, in superconductors at $T<T_{cr}$. Those cases are called the spontaneous symmetry breaking in physical systems. At $T>T_{cr}$ the symmetry recovers, superconductivity vanishes.

Note the draw ratio at break falls with the molecular weight increase following the power law with the exponent equal to $-0.4 - -0.5$ in dependence on the drawing temperature [28-31]. Thus, the neck draw ratio is more convenient parameter for studying.

Let us recall that we define the ratio $(W/W_{cr})^{-2}$ as a probability (see [11]). We can consider the value as the square of a probability and regard $W/W_{cr} = \lambda_n$ as the reciprocal probability which may be connected with the neck draw ratio [21] taking into account reasons mentioned above.

Another sense can be attributed to the critical point $N_{cr}$. According to [11]

$$\lambda_n = (C/B)^{1/2} N^{-\beta} \tag{32}$$

where $B$ and $C^2 = N_{cr}$ are constant. Assuming $\lambda_n (\lambda_n^{\perp})^2 = 1$, $(\lambda_n^{\perp})^2$ is the ratio of the final cross-section of the sample to its initial, that is, the volume does not change during drawing, one can find

$$\lambda_n^{\perp} = B^{1/4} N^{\beta/2} C^{-1/4}$$

and obtain about $N = N_{cr}$ at $C^2 = N_{cr}$

$$\lambda_n^{\perp} = const\ N_{cr}^{-1/8}.$$

The law is similar to one in the 2D Onsager problem with $\beta = 0.125$ (15). Thus, several critical points can be observed at $N = N_{cr}$.

Let us assume that the stress is the following function of time:

$$S(t) = S\ exp(-t/t_M)$$

and

$$\partial S(t)/\partial t = - S/t_M \tag{33}$$

where $S = \sigma a^3/T$ is the dimensionless (quasi)equilibrium value while $\sigma$ is the true stress applied to a polymeric sample, $t_M$ is the Maxwell relaxation time.

Let us determine the quadratic form characterized the energy dissipation [9] upon the plastic deformation of polymers

$$f = 0.5 \, (S/t_M) \, (\lambda_n - 1)^2. \tag{34}$$

The variation of the stress can be written from (33) and (34) as

$$\partial S(t)/\partial t = - (\lambda_n - 1)^{-1} \partial f/\partial (\lambda_n - 1) = - S/t_M.$$

The variation entropy rate, $ds/dt$, is defined by the following equations [9]

$$ds/dt = (\partial s/\partial S(t))(\partial S(t)/\partial t) = - (\lambda_n - 1)^2 \, (\partial S(t)/\partial t) = 2f. \tag{35}$$

Thus, at the isotropic state as well as at the critical point $\lambda_n = 1$ [11] then from (35) $ds/dt = 0$. The entropy has a maximum at the equilibrium state. There is no drawing and, consequently, there is no dissipation of energy. The macroscopic consideration can serve as an model of the neck formation under irreversible deformations of polymers. At the critical point nothing change, but if $N \neq N_{cr}$ upon acting the stress the spontaneous symmetry breaking occurs and the polymer sample deforms through the neck formation.

The elastic modulus, $E$, can be expressed as the derivative of the thermodynamic potential with respect to the stress. Put another way, one can write using (35)

$$E = -(ds/dS) \approx -(ds/dt)/(\partial S(t)/\partial t) \sim (\lambda_n - 1)^2.$$

Upon the neck formation let us assume from (34)

$$S \to S \, (\lambda_n - 1), \, t_M \to t_M \, / \, (\lambda_n - 1).$$

Then

$$\eta \sim E \, t_M \sim (\lambda_n - 1),$$

$$v_{dr} \sim S/\eta = const$$

and one can see that the stationary state [16] with draw velocity $v_{dr} = const$ is observed. Here, $\eta$ is the effective viscosity of the system. The minimum of the energy dissipation takes place during drawing [16].

2 8 A simple example of solving problems.

Let us recall the percolation scaling theory in gels [8, 17]. If we know an exponent from precise experiments then we can easy find the rest exponents. Let us suppose $n = 6$ then about the gel point the correlation length exponent $\nu = 0.87$, the weight average polymerisation degree exponent $\gamma = 1.71$ and the gel fraction exponent $\beta = 0.45$. The exponent estimated from (16) and (18) with the use of (26) and (30) can be a good approximation at initial stages of researches.

3 CONCLUSION REMARKS

In the paper we have shown that the high energy part of the elementary particle spectrum (the gauge fields) can control the low energy dissipation processes through the mechanism of the spontaneous symmetry breaking. Our current and previous theoretical results [11, 21] with experimental data [21, 28-31] confirm it.

**Acknowledgements**

The author is thankful to Russian Foundation for Basic Research (Grant No 01-03-32225) for financial support, to Professor S. N. Chvalun from Moscow Institute of Physics and Technology for helpful discussions, and to Dr. A. V. Mironov for technical help.